\documentclass[journal]{IEEEtran}
\usepackage{cite}

\usepackage{graphicx}
\usepackage[flushmargin]{footmisc}
\usepackage{listings}
\usepackage{amsmath,amssymb,amsfonts}
\usepackage{float}
\usepackage{hyperref}
\hypersetup{colorlinks,allcolors=black}
\usepackage[skip=10pt]{caption}
\usepackage{bm}
\usepackage[inline]{enumitem}
\usepackage{soul}
\usepackage{multirow}
\usepackage{multicol}
\usepackage{setspace}
\usepackage{tabularx}

\ifCLASSINFOpdf
\else
\fi

\usepackage{subfig}

\begin{document}
\title{Deep Adversarial Learning on Google Home devices}

\author{Andrea Ranieri${}^{1}$, Davide Caputo${}^{2}$, Luca Verderame${}^{2}$, Alessio Merlo${}^{2}$, and Luca Caviglione${}^{1}$\\
${}^{1}$IMATI - National Research Council of Italy\\{\{andrea.ranieri, luca.caviglione\}@ge.imati.cnr.it}\\
${}^{2}$DIBRIS - University of Genova, Italy\\
{\{davide.caputo, luca.verderame, alessio\}@dibris.unige.it}}


%
%

\markboth{IEEE COMMUNICATIONS LETTERS, VOL. $X$, NO. $Y$, $MONTH$ 2020}%
{Ranieri \MakeLowercase{\textit{et al.}}: Deep Adversarial Learning on Google Home devices}

\maketitle

\begin{abstract}

Smart speakers and voice-based virtual assistants are core components for the success of the IoT paradigm. Unfortunately, they are vulnerable to various privacy threats exploiting machine learning to analyze the generated encrypted traffic. To cope with that, deep adversarial learning approaches can be used to build black-box countermeasures altering the network traffic (e.g., via packet padding) and its statistical information.
This letter showcases the inadequacy of such countermeasures against machine learning attacks with a dedicated experimental campaign on a real network dataset. Results indicate the need for a major re-engineering to guarantee the suitable protection of commercially available smart speakers.

\end{abstract}


\IEEEpeerreviewmaketitle

\section{Introduction}
\label{sec:introduction}

The popularity of smart speakers (including voice-based virtual assistants) is rooted in their ability to control IoT nodes, network appliances, and other devices via natural speech.
They can also be used to access multimedia contents and obtain various information, including news, weather forecasts, and traffic conditions. 
To implement such functionalities, the speaker exchanges information with a remote data center, leading to several security issues, e.g., device enumeration attacks, mass profiling, and privacy threats. An emerging trend in cybersecurity exploits machine learning techniques to obtain information from the encrypted traffic exchanged by the speaker with its ecosystem  \cite{iotsecsur, Apthorpe2017c}.

Literature abounds in works investigating how statistical analysis of network flows produced by smart devices can be abused for reconnaissance or attack purposes. For instance, the traffic produced by home devices can be used to understand if a user is at home \cite{Copos2016} as well as to model daily routines \cite{Acar2018} or the sleep cycle \cite{Apthorpe2017c}. 
In general, attacks leveraging machine learning proved to be effective, even when relying upon ``poor'' information. As an example, IoT nodes and connected devices can be identified by simply using the length of the produced protocol data units \cite{trapacklen}. When HTTP-based interactions are present, it is possible to infer precise details, e.g., the status of a light bulb, as well as hijacking the conversation or physically endanger the target \cite{Amar2018}.

An emerging research trend explores the use of various artificial intelligence and machine learning techniques to classify the commands issued to smart speakers (see, e.g., \cite{caputo2020fine} and the references therein). 
To this aim, attackers take advantage of traffic features not protected by the encryption, such as inter-packet time, throughput, the location of some endpoints, and the number of connections.
Since the classification is typically accurate, the attacker can infer details like the number of devices controlled by the smart speaker, the presence of the user (even when the interaction is absent), and the ``kind" of the issued commands \cite{caputo2020fine, Acar2018,Amar2018,Copos2016, trapacklen}. 
Moreover, a relevant part of the traffic produced by smart speakers shares functional and technological traits with VoIP, meaning that it is also susceptible to attacks for disclosing the language of the talker or other sensitive behaviors \cite{voiplanguage}.

Therefore, this letter focuses on investigating deep adversarial learning countermeasures against machine learning attacks targeting the traffic produced by smart speakers. 
To the best of our knowledge, this aspect has been mostly overlooked so far.
The only notable exception is \cite{adaptive}, which proposes a padding scheme to protect IoT and smart devices from statistical analysis. Instead, our work aims to showcase the limitations of traffic manipulation or morphing approaches, which often lead to flawed countermeasures \cite{parrot}. 

To do so, we built a new dataset containing the network traffic of a typical smart home environment and an experimental testbed to evaluate the efficacy of deep adversarial learning techniques.
The experimental activities exploit both theoretical approaches, i.e., the usage of Savitzky–Golay filters and Additive White Gaussian Noise (AWGN) on all the statistical features, and realistic ones, named ``\textit{Realistic Adversarial}", that use constant padding and AWGN techniques that considers the constraints of the protocols and networks in use.
The achieved results argue that smart speaker privacy needs a complete rethink.




The rest of the letter is structured as follows: Section \ref{sec:background} provides the background and the threat model, Section \ref{sec:adversarial-setup-2} presents the deep adversarial techniques used in this work. Section \ref{sec:testbed} describes the evaluation testbed and Section \ref{sec:results} presents the obtained results. Finally, Section \ref{sec:conclusions} concludes the letter. 

\section{Background and Attack Model}
\label{sec:background}

\begin{figure}[h]
    \centering
    \includegraphics[width=0.45\textwidth]{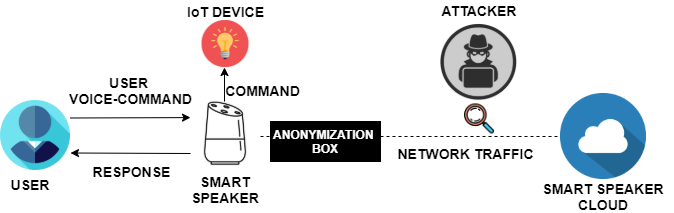}
    \caption{MITM threat model of a smart home ecosystem.}
    \label{fig:defense_model}
\end{figure}

Figure \ref{fig:defense_model} depicts a typical smart speaker ecosystem that provides the voice-activated user interface and acts as a hub for other IoT nodes and network appliances.
In essence, the speaker collects, samples, and transmits voice commands to remote cloud services in charge of processing data to deliver back textual/binary representations as well as additional content, e.g., multimedia streams. 
The smart speaker can also provide feedback to the user, play content, retrieve data from third-party providers (e.g., music streaming services) or drive other nodes via local network or short-range links like IEEE 802.15.4. 
Even if the information gathered by commands exchanged locally between the speaker and various nodes can be used to threaten the privacy of the ecosystem \cite{nour2019security}.
This letter focuses on the attacks exploiting the network traffic exchanged between the speaker and its remote cloud. 

From a security perspective, the continuous exchange of data between the smart speaker and the cloud is a prime point of fragility. As depicted in Figure \ref{fig:defense_model}, an adversary (denoted as \textit{attacker}) can mount MITM (Man-in-the-Middle) attacks \cite{Andy2017MQTT} to gather network traffic even in the case of a communication encrypted with TLS/SSL \cite{sslsurvey}.

Even if many commercial smart speakers implement countermeasures to protect the network traffic, the majority is still prone to a variety of privacy-breaking attacks targeting a composite set of features observable within the encrypted traffic flows \cite{monkey,iotsecsur}. Specifically, we  focus on an attacker willing to use machine learning or deep learning algorithms on encrypted traffic samples to infer ``behavioral'' information, e.g., the presence of the victim or the ``type'' of the requested information \cite{caputo2020fine, Acar2018,Amar2018,Copos2016, trapacklen}. Owing to the end-to-end encryption, the attacker can only observe and acquire the traffic produced by the smart speaker and cannot alter, manipulate or perform deep packet inspection operations. The attacker can then only rely on general statistics, e.g., the  throughput, the size of protocol data units, IP addresses, the number of different endpoints, flags within the headers of the packets, or the behavior of the congestion control of the TCP \cite{caputo2020fine}.

The standard approach to mitigate machine learning attacks on the network traffic exploits the use of a middlebox (denoted as \textit{anonymization box} in Figure \ref{fig:defense_model}) able to ``sanitize'' the network traffic by removing (or altering) the data that the attacker can exploit. 
For instance, the anonymization box can pad packets  \cite{adaptive} or perform NAT-like operations to prevent profiling endpoints or probing \cite{gregorczyk2020sniffing}. 
All in all,  since the anonymization box is outside the device, it cannot alter the protocol/communication architecture of the smart speaker ecosystem. 
Rather, it can only manipulate the traffic without disrupting the flow or penalizing the QoE perceived by the user, for instance, in terms of real-time guarantees (see, e.g., \cite{fan2017spabox} and references therein). In the following, we will showcase the limits of such an approach, which appears to be unsuited to face modern machine learning-capable threats.

\section{Deep Adversarial Learning Techniques}\label{sec:adversarial-setup-2}

This letter investigates both theoretical and practical deep adversarial learning techniques to lower the classification accuracy.
Ideally, we would like to reduce the classification accuracy to be as close as possible to a ``coin toss" (e.g., $50\%$ on a two-class classification problem).
As in previous works \cite{caputo2020fine,su2020you,Shahid2019}, the attacker can only acquire encrypted traffic to compute statistical metrics and analyze them using machine learning techniques. Such a computation requires using a suitable number of packets grouped using either time spans of length $\Delta t$ or bursts of a fixed size of $N$ packets.

The methods considered for this work are: \textit{i}) smoothing of features through Savitzky--Golay filter, \textit{ii}) injecting Additive White Gaussian Noise (AWGN) into the features time series and \textit{iii}) applying a \textit{Realistic Adversarial}, i.e., a targeted approach to feature degradation that also considers the constraints of the protocols and networks in use.

The first two approaches aim to show the theoretical performances that could be obtained by randomizing, without any constraints, the statistics of the packets and the features derived from them.
In detail, the Savitzky--Golay filter \cite{savitzky1964smoothing} allows smoothing all features through a moving window. The length of the filter window determines the number of samples taken into consideration, and a polynomial of user-defined order subsequently approximates these samples.

The second adversarial technique uses Additive White Gaussian Noise (AWGN) -- therefore with zero mean -- whose variance is set proportionally to the variance of the original signal subject to adversarial. 

The Realistic Adversarial, instead, represents an approach that takes into account which features can actually be distorted at the egress of the IoT device (therefore with external hardware such as an \textit{anonymization box}) without compromising its operation (e.g., it is possible to add padding to the packets, but it is not possible to randomize the TCP window without disrupting the service completely).
In this work, we exploited two Realistic Adversarial techniques: the \textit{first} is a  \textbf{constant padding} within the whole time series, which was simulated selecting the maximum value of the mean TCP packet length (\textit{mean\_len\_pack}). 
The standard deviation of the TCP packet length (\textit{std\_len\_pack}) was set to zero to match this operation.
The \textit{second} technique concerns the injection of \textbf{AWGN} in the following features:
\begin{itemize}
  \item \textit{std\_ipt}: to simulate jitter while sending packets;
  \item \textit{n\_pack\_tcp} and \textit{n\_pack\_udp}: to simulate decoy connections and packets between endpoints;
  \item \textit{n\_pack\_icmp}: to simulate decoy ping/traceroute packets between endpoints;
  \item \textit{n\_port\_unique}:, to simulate decoy TCP/UDP packets addressed to random port numbers.
\end{itemize}
Moreover, to match the realistic scenario, we cannot modify the following features: \textit{max\_diff\_time}, \textit{n\_ip\_unique}, \textit{mean\_window} and \textit{std\_window}.

\begin{table*}[h]
    \renewcommand\arraystretch{1.35}
    \scriptsize
    \centering
    \begin{tabular}{|l|c||l|c|}
        \hline
        \textbf{Feature Name} &\textbf{Adversarial Technique} & \textbf{Feature Name} &\textbf{Adversarial Technique} \\
        \hline \hline
       
        Number of different IP address (\textit{n\_ip\_unique}) & None &  Number of different TCP/UDP ports (\textit{n\_port\_unique}) & AWGN\\ \hline
        
        Number of TCP packets (\textit{n\_pack\_udp}) & AWGN &  Number of UDP packets (\textit{n\_pack\_udp}) & AWGN \\\hline 
        
        Number of ICMP packets (\textit{n\_pack\_icmp}) & AWGN & \textit{Per-}window Inter packet time (\textit{max\_diff\_time}) & None \\\hline

        Average of TCP Window (\textit{mean\_window})  & None & Standard deviation of TCP Window (\textit{std\_window}) & None \\\hline
       
        Average of IPT (\textit{mean\_ipt}) & None & Standard deviation of IPT (\textit{std\_ipt})  & AWGN\\ \hline
        
        Average of packet length (\textit{mean\_len\_pack}) & Const. Padding & Standard Deviation of packet length (\textit{std\_len\_pack}) & Set to Zero\\ \hline

         
    \end{tabular}
    \caption{Name and Acronym of statistical indicator used and the adversarial techniques applied to them}
    \label{tab:features_name_statistical_indicator}
\end{table*}


\section{Experimental Testbed}
\label{sec:testbed}

We developed an experimental testbed to prove the effectiveness of privacy threats and the inadequacy of countermeasures applied to smart speakers' network traffic. 
Briefly, we collected a dataset with the network traffic of the IoT device during typical usage scenarios, e.g., during voice queries or media playback. 
Then, we used a set of deep adversarial learning techniques to alter the statistical information of outbound traffic. 
Finally, we applied a set of ML techniques to evaluate the corresponding degradation of the classification process.
The experimental activities were carried out on an Intel Core i7-3770 computer equipped with $16$ GB of RAM and Ubuntu 16.04 LTS, and a Google Home Mini smart speaker. 
The traffic has been captured via an instrumented computer acting as an IEEE 802.11 access point running ad-hoc scripts for {\tt tshark}\footnote{\url{https://www.wireshark.org/docs/man-pages/tshark.html}}.
All the generated traffic traces have been anonymized and made available through Kaggle\footnote{\url{https://www.kaggle.com/smartspeaker/google-home-pcap}}.

\subsection{Dataset Definition}
In this letter, we extend the dataset already published in our previous work \cite{caputo2020fine}. 
Briefly, the original dataset consists of 9-days of network traffic that comprise: i) traffic with the microphone disabled (\textit{Mic On-Off}), ii) microphone enabled in a quiet environment, and iii) microphone enabled with background noise (\textit{Mic On-Noise}).

Thus, to mimic the normal use of smart speakers by users, we extended the available data with three different classes of queries for the smart speaker, i.e., media, travel, and utility\footnote{\url{https://voicebot.ai/2019/03/12/smart-speaker-owners-agree-that-questions-music-and-weather-are-killer-apps-what-comes-next/}}.
To do so, we executed three different rounds of measurements that last three days each.
In essence, a synthetic talker has been created by using various voice records representing a wide range of speakers (e.g., male and female or with different accents or talking speeds) and it has been used to issue commands to the smart speaker.
In the first round, we focused on retrieving the network traffic generated to playback multimedia content. 
For example, we captured traffic when the synthetic talker asked questions like \textit{``What's the latest news?"} or \textit{``Play some music"}.
For the second round, we performed queries related to travels, thus accounting for the interaction with services providing traffic indications or weather forecasts.
In this case, we asked questions like \textit{``How is the weather today?"}. Lastly, we performed general queries belonging to the utility category, like \textit{``What's on my agenda today?"} and \textit{``What time is it?"}.

For each query, we collected $30$ seconds of inbound and outbound network traffic to have a proper tradeoff between accuracy and size of the data. 
The newly collected data contains $2,500k$  packets for media queries, $400k$ packets for travel queries, and $310k$ for utility queries.

\subsection{Deep Adversarial Techniques Setup}

We implemented the three deep adversarial learning techniques presented in Sect. \ref{sec:adversarial-setup-2}. 
In detail, we set the Savitzky–Golay filter moving window to 51, and we applied different polynomials degrees, i.e., $\psi=\{1, 3, 5, 7, 9\}$.
For the Additive White Gaussian Noise (AWGN) techniques, we multiplied the original variance using a constant $\nu=[0.2, 2.0]$ for the \textit{Mic On-Off} and \textit{Mic On-Noise} scenario and $\nu=\{2, 8, 16, 32, 64\}$ for the \textit{utility/media/travel} scenario.

Table \ref{tab:features_name_statistical_indicator} lists the set of network features computed for the experimental activity of this letter and the corresponding employed deep adversarial technique.

\subsection{ML techniques}

To implement the attacker, we used machine learning algorithms provided by the \textit{scikit-learn} and \textit{Fast.ai} libraries. 
We considered the most popular techniques commonly used in the literature, i.e., AdaBoost (AB), Decision Tree (DT), k-Nearest Neighbors (kNN), Random Forest (RF) and Neural Networks (NN) \cite{caputo2020fine,li2007accurate,yang2008p2p}.
In our experiment, we assume that the attacker is able to collect 3 days of traffic for each scenario, similar to \cite{caputo2020fine}. 
The classifiers were neither pre-trained with the original data nor fed with previously trained models.


\begin{figure*}[h]
\centering


\subfloat[Mic On-Off Adversarial with smoothing]{\includegraphics[width=0.33 \linewidth]{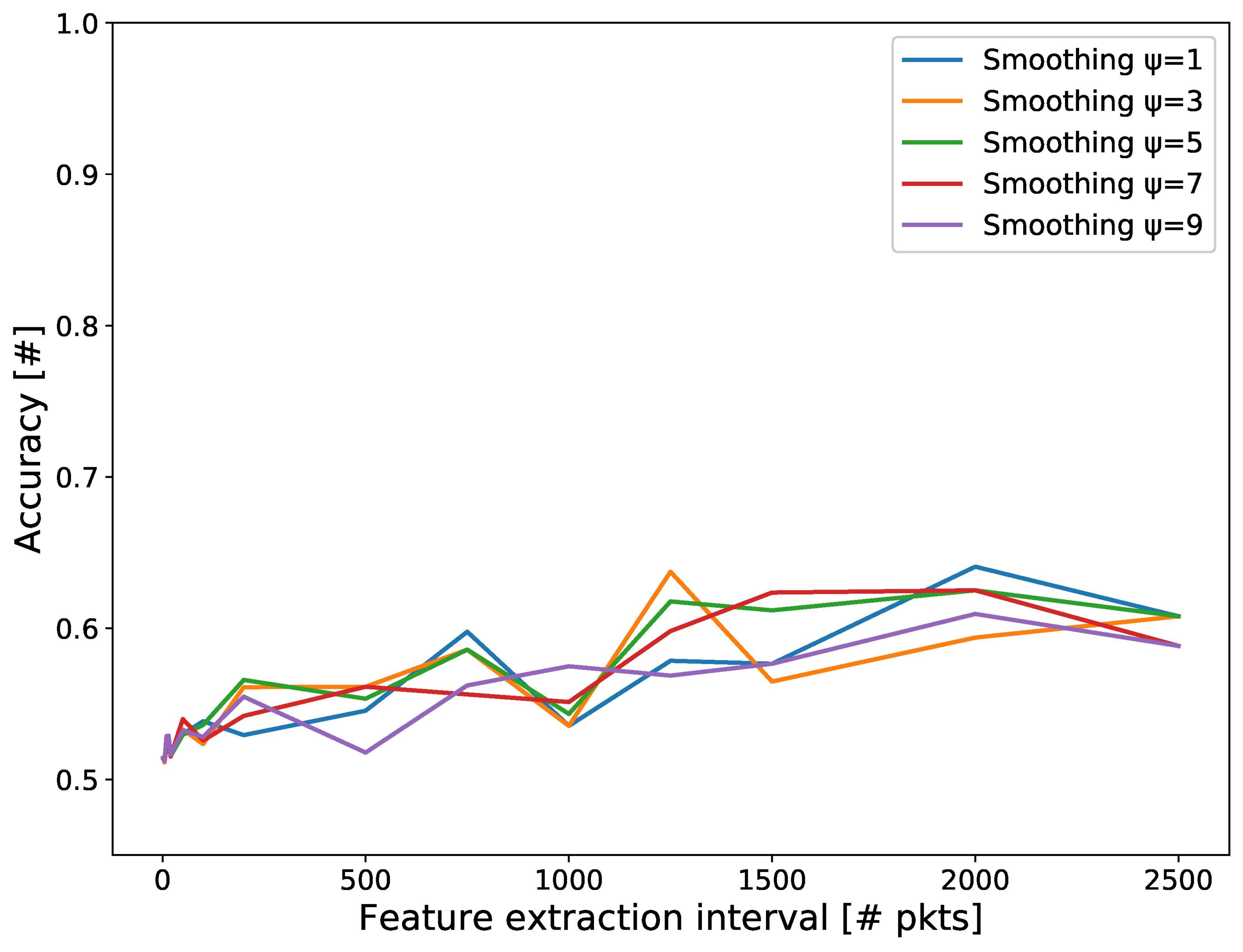}\label{fig:on-off-smoothing}}
\subfloat[Mic On-Off Adversarial with noise]{\includegraphics[width=0.33 \linewidth]{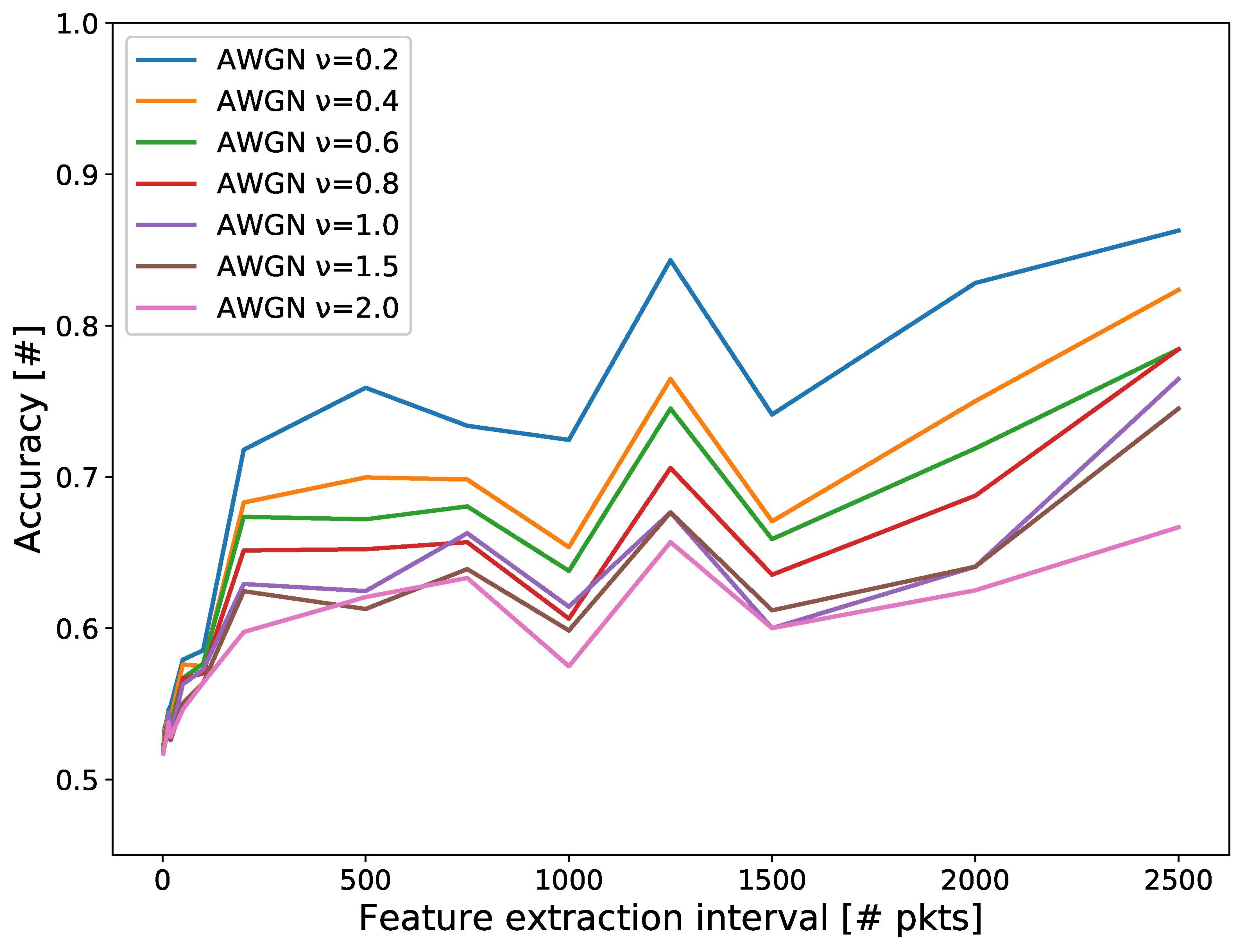}\label{fig:on-off-noise}}
\subfloat[Mic On-Off Realistic adversarial]{\includegraphics[width=0.33 \linewidth]{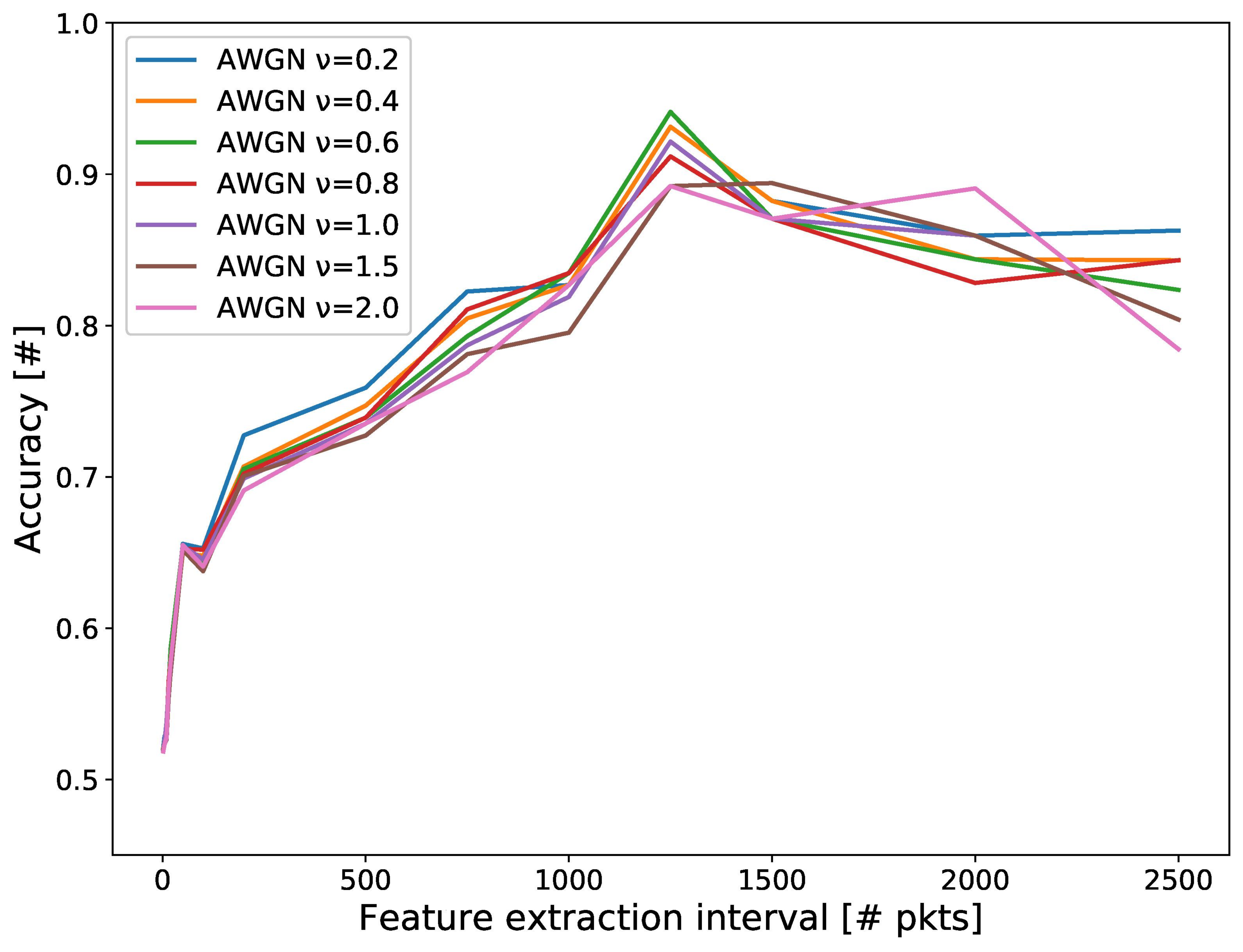}\label{fig:on-off-realistic}}

\subfloat[Mic On-Noise Adversarial with smoothing]{\includegraphics[width=0.33 \linewidth]{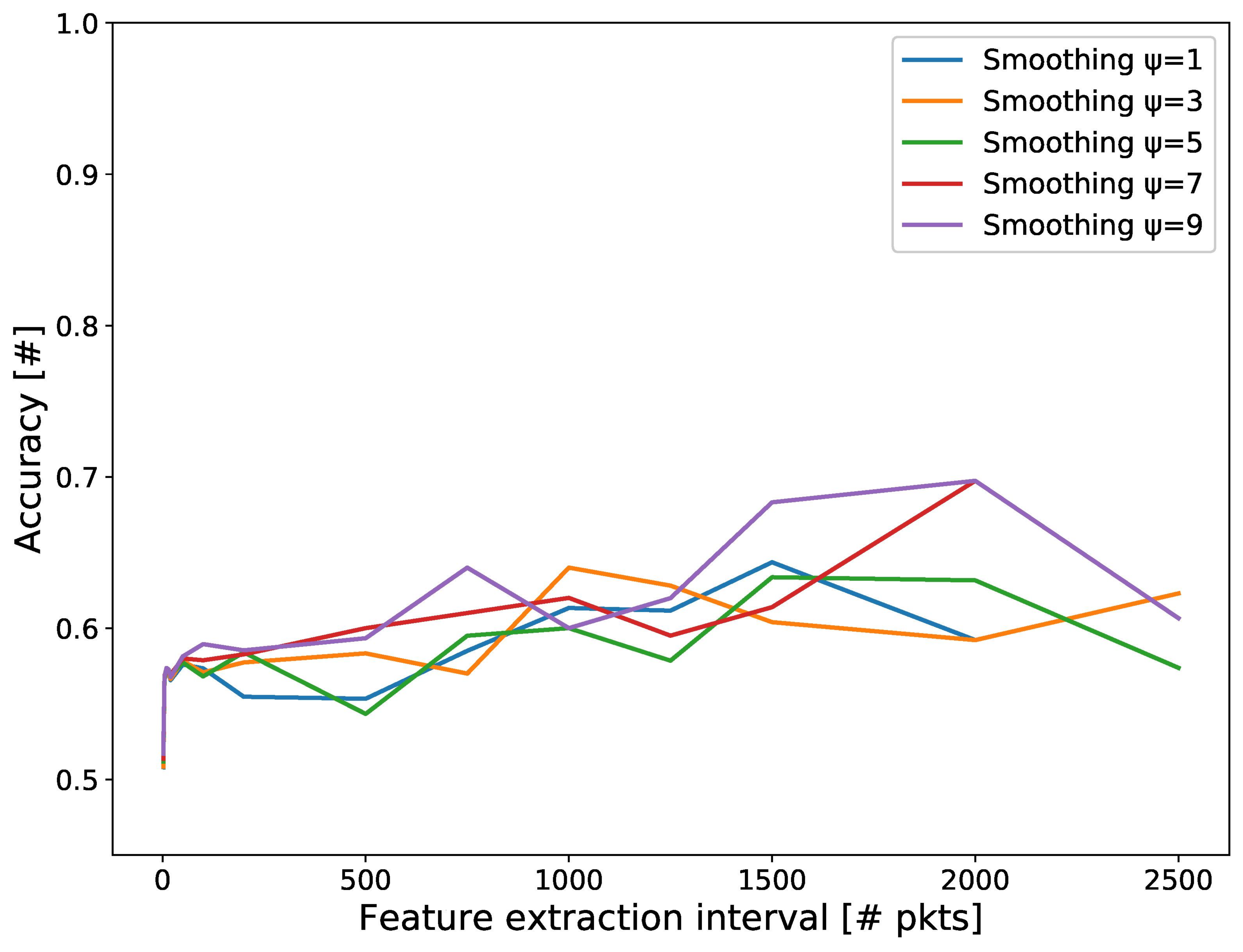}\label{fig:on-noise-smoothing}}
\subfloat[Mic On-Noise Adversarial with noise]{\includegraphics[width=0.33 \linewidth]{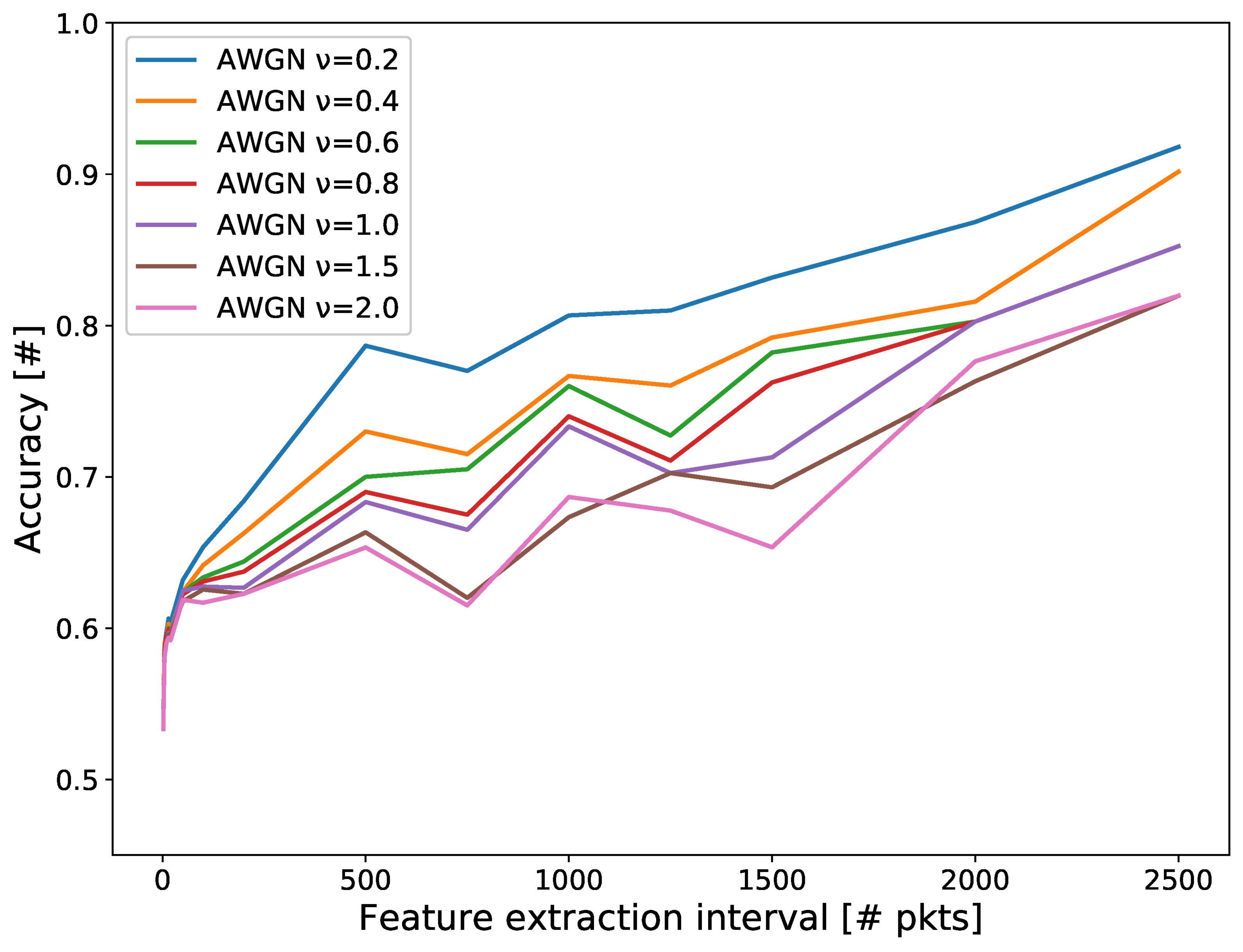}\label{fig:on-noise-noise}}
\subfloat[Mic On-Noise Realistic adversarial]{\includegraphics[width=0.33 \linewidth]{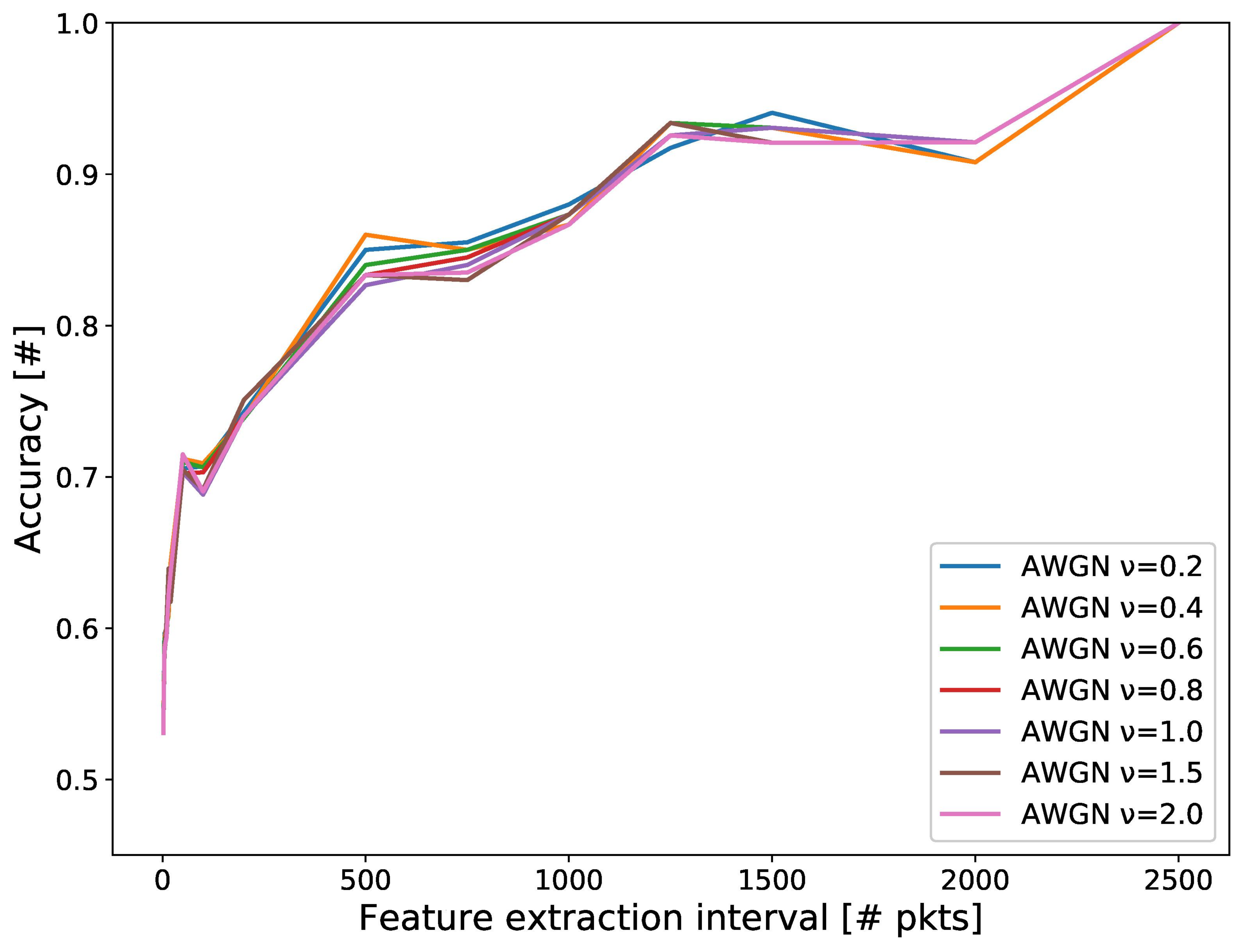}\label{fig:on-noise-realistic}}

        \caption{\textit{Mic On/Off On/Noise scenario}: classification accuracy of the neural network model after different adversarial techniques have been applied to the original features: (a),(d) \textit{Savitzky--Golay} filter, (b),(e) \textit{AWGN}, (c),(f) \textit{Realistic adversarial technique.}}
        \label{fig:classification_adversarial_mic_on_off_noise_pkts}
\end{figure*}

\begin{figure*}[h]
\centering
\subfloat[\textit{Utility/Media/Travel} Adversarial with smoothing]{\includegraphics[width=0.33 \linewidth]{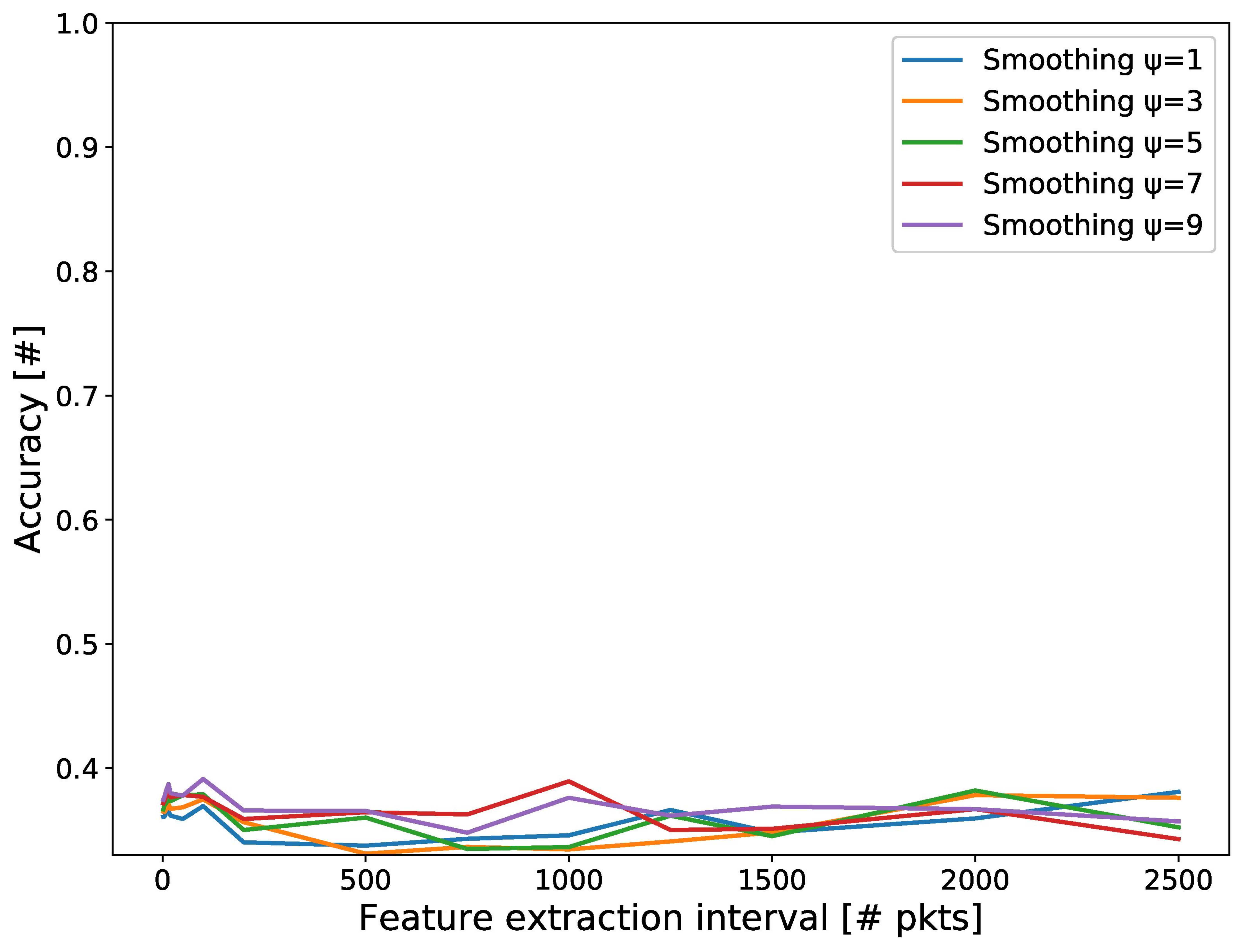}\label{fig:time-news-meteo-smoothing}}
\subfloat[\textit{Utility/Media/Travel} Adversarial with noise]{\includegraphics[width=0.33 \linewidth]{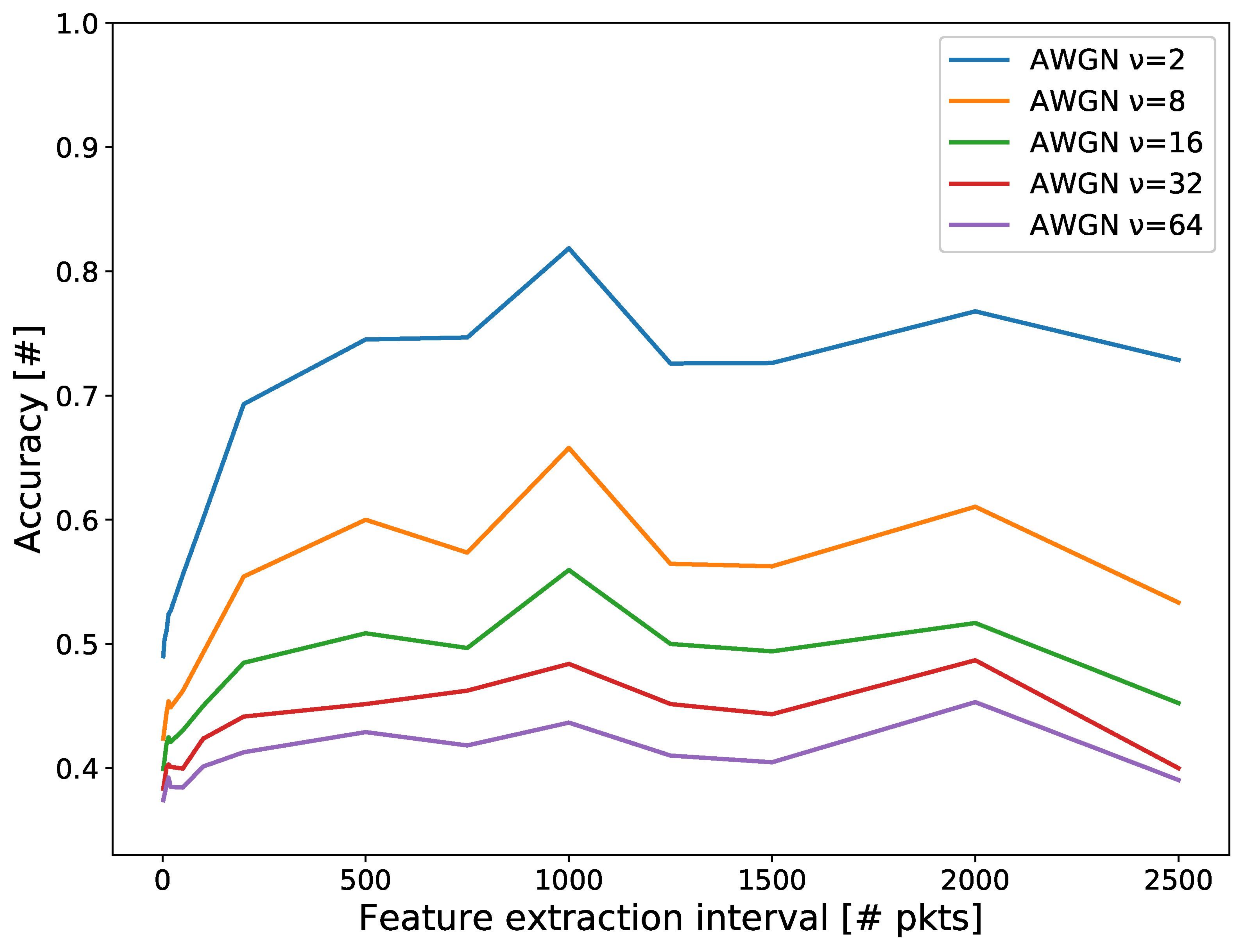}\label{fig:time-news-meteo-noise}}
\subfloat[\textit{Utility/Media/Travel} Realistic adversarial]{\includegraphics[width=0.33 \linewidth]{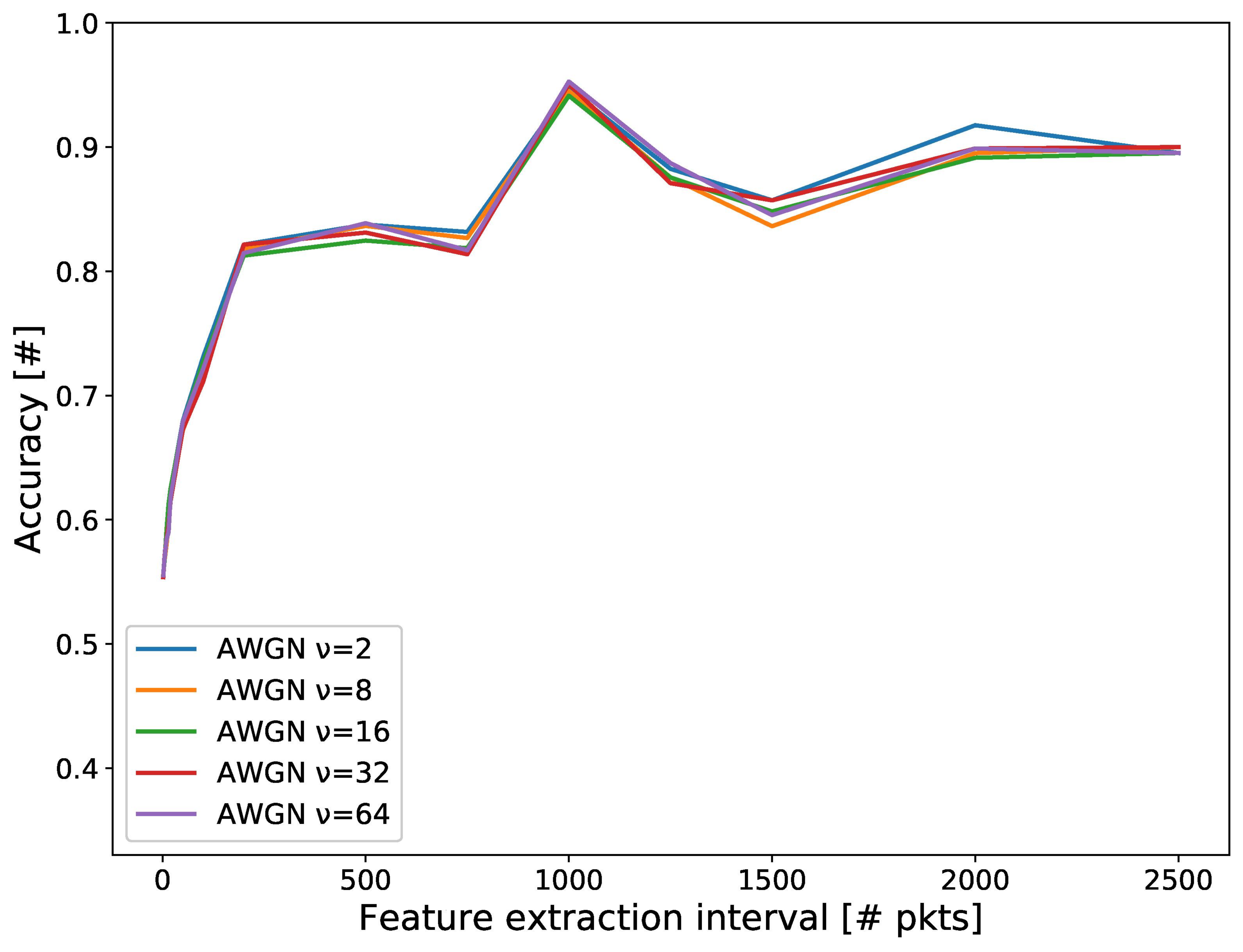}\label{fig:time-news-meteo-realistic}}

        \caption{\textit{Utility/Media/Travel scenario}: classification accuracy of the neural network model after different adversarial techniques have been applied to the original features: (a) \textit{Savitzky--Golay} filter, (b) \textit{AWGN}, (c) \textit{Realistic adversarial technique.}}
        \label{fig:classification_adversarial_time_news_meteo_pkts}
\end{figure*}


\section{Experimental Results}
\label{sec:results}
In this section, we show and discuss the numerical results obtained. 
First, we show the performance and accuracy of machine learning algorithms used to infer the query category. 
Next, we discuss the efficacy of the deep adversarial learning techniques to protect the network traffic. 

\begin{figure}[h]
    \centering
    \includegraphics[width=0.4\textwidth]{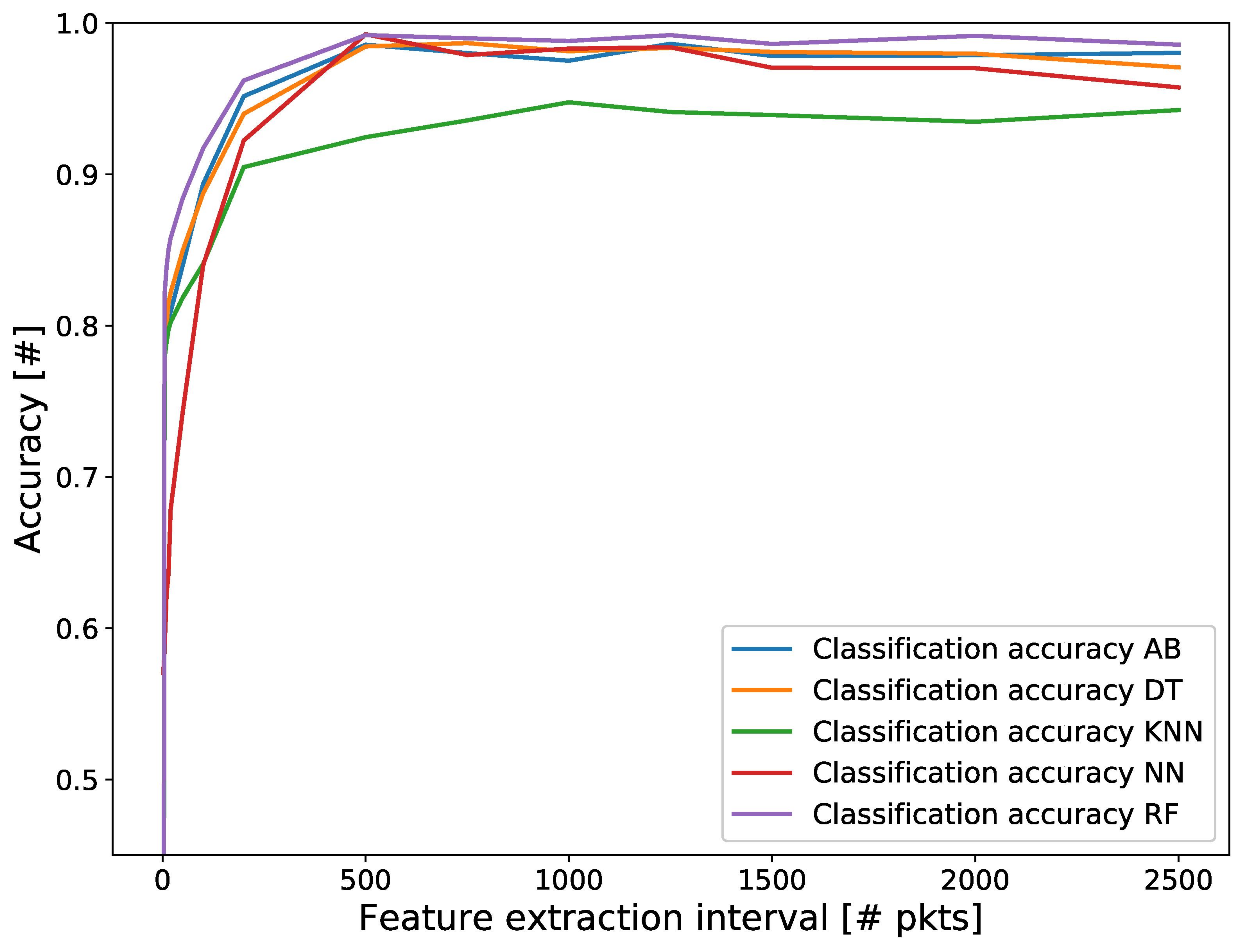}
    \caption{\textit{Utility/Media/Travel scenario}: classification accuracy of the different machine learning/deep learning models using unmodified original features.}
    \label{fig:plain_classification_time_news_meteo_pkts}
\end{figure}

\textbf{Query Classification.}
\figurename\textit{~\ref{fig:plain_classification_time_news_meteo_pkts}} shows the classification accuracy for ML techniques trained on the original data (not subjected to any adversarial technique) of the \textit{Utility/Media/Travel} scenario. 
With sampling and feature generation intervals higher than $500$ packets, all the considered algorithms achieve a classification accuracy higher than $90\%$, being the kNN algorithm the less precise. 
The accuracy of the neural network model (i.e., NN) also tends to drop slightly as the interval increases because the number of samples decreases too much for the training ``from scratch" of a neural network (at $1500$ packets, the number of samples is just above $300$ units for this scenario).
Such behavior is also consistent with the classification of the \textit{Mic On-Off} and \textit{Mic On-Noise} scenarios whose accuracy is higher than $80\%$ (as described in \cite{caputo2020fine}).

\textbf{Mitigation Results.}
The set of images in \figurename\textit{~\ref{fig:classification_adversarial_mic_on_off_noise_pkts}}  and \figurename\textit{~\ref{fig:classification_adversarial_time_news_meteo_pkts}} depict the accuracy of a classifier trained on data subjected to the three deep adversarial techniques.

In detail, \figurename\textit{~\ref{fig:on-off-smoothing}} to \textit{\ref{fig:on-off-realistic}} show the accuracy obtained in the \textit{Mic On-Off} classification scenario, \figurename\textit{~\ref{fig:on-noise-smoothing}} to \textit{\ref{fig:on-noise-realistic}} refer to the \textit{Mic On-Noise} scenario, and \figurename\textit{~\ref{fig:classification_adversarial_time_news_meteo_pkts}} shows the classification performance of a classifier trained on the traffic features from the \textit{Utility/Media/Travel} scenario.

In the first two scenarios, the trend is similar: the theoretically most effective adversarial technique is the polynomial smoothing as it deprives the signal of most of its high-frequency components. 
On the contrary, AWGN injection is less effective as it leaves much of the information untouched, especially with low values of $\nu$ (i.e. variance multiplier values).
\figurename\textit{~\ref{fig:on-off-realistic}} and \textit{\ref{fig:on-noise-realistic}}, conversely, show the implementation of a realistic adversarial technique which is therefore not able to degrade all the features simultaneously. 
As can be seen from the accuracy levels (for some sampling intervals $> 1250$ packets, higher than $90\%$), the neural network model is sufficiently ``intelligent" to correctly predict the class using just the few remaining features and ignoring all the others. 
In this case, higher AWGN intensity does not appear to affect the prediction in any way.

The analysis of the queries scenario in \figurename\textit{~\ref{fig:classification_adversarial_time_news_meteo_pkts}} confirms that the polynomial smoothing techniques are the most efficient, bringing the accuracy of the classifier to around $40\%$, very close to the theoretical level of a 33\% random choice of a problem with three classes (cf. \figurename\textit{~\ref{fig:time-news-meteo-smoothing}}). 
On the contrary, \figurename\textit{~\ref{fig:time-news-meteo-noise}} highlights the need for a substantially higher $\nu$ value for the AWGN in order to sufficiently degrade the features for this scenario. Indeed, unlike the \textit{Mic On-Off} and \textit{Mic On-Noise} scenarios (with $\nu=2.0$), this scenario requires a variance multiplier $\nu = 64$ to bring the accuracy level of the ML classifiers to values close to the $33\%$.

Finally, \figurename\textit{~\ref{fig:time-news-meteo-realistic}} shows the accuracy obtained from the same neural network architecture trained on degraded features with realistic adversarial. 
As in \figurename\textit{~\ref{fig:on-off-realistic}} and \textit{\ref{fig:on-noise-realistic}}, the impossibility of degrading all the features at once, leaving some of them intact, preserves enough information for the neural network to learn how to correctly identify the three classes of the problem, even with an accuracy of $95\%$ considering an interval of $1000$ packets, independently from the magnitude of the injected AWGN.




\section{Conclusions}
\label{sec:conclusions}

In this letter we have empirically demonstrated how adversarial learning countermeasures applied to the virtual assistant's outbound traffic are ineffective against machine learning attacks, thus leading to serious concerns for the privacy of users in smart home environments.

The results indicate the need for a major HW/SW redesign of virtual assistant platforms to ensure adequate protection of commercially available smart speakers.

\ifCLASSOPTIONcaptionsoff
  \newpage
\fi


\bibliographystyle{IEEEtran}
\bibliography{biblio}

%




\end{document}